\documentclass[aps,prx,twocolumn,superscriptaddress]{revtex4-2}
\usepackage{amssymb,amsbsy,graphicx,xcolor,times,subfigure,marvosym,color,bm,multirow,epstopdf}
\usepackage[latin1]{inputenc}
\begin{document}

\title{Fractional magnetoresistance oscillations in spin-triplet superconducting rings}

\author{G\'abor~B.~Hal\'asz}
\affiliation{Materials Science and Technology Division, Oak Ridge National Laboratory, Oak Ridge, Tennessee 37831, USA}
\email{This manuscript has been authored by UT-Battelle, LLC under Contract No. DE-AC05-00OR22725 with the U.S. Department of Energy. The United States Government retains and the publisher, by accepting the article for publication, acknowledges that the United States Government retains a non-exclusive, paid-up, irrevocable, world-wide license to publish or reproduce the published form of this manuscript, or allow others to do so, for United States Government purposes. The Department of Energy will provide public access to these results of federally sponsored research in accordance with the DOE Public Access Plan (http://energy.gov/downloads/doe-public-access-plan).}

\begin{abstract}

\textbf{Half-quantum vortices in spin-triplet superconductors are predicted to host Majorana zero modes and may provide a viable platform for topological quantum computation. Recent works also suggested that, in thin mesoscopic rings, the superconducting pairing symmetry can be probed via Little-Parks-like magnetoresistance oscillations of periodicity $\Phi_0 = h / 2e$ that persist below the critical temperature. Here we use the London limit of Ginzburg-Landau theory to study these magnetoresistance oscillations resulting from thermal vortex tunneling in spin-triplet superconducting rings. For a range of temperatures in the presence of disorder, we find novel oscillations with an emergent fractional periodicity $\Phi_0 / n$, where the integer $n \geq 3$ is entirely determined by the ratio of the spin and charge superfluid densities. These fractional oscillations can unambiguously confirm the spin-triplet nature of superconductivity and directly reveal the tunneling of half-quantum vortices in candidate materials such as Sr$_2$RuO$_4$ and UTe$_2$.}

\end{abstract}

\maketitle

\textbf{Introduction.} Flux quantization is a defining feature of superconductivity that directly originates from the macroscopic quantum coherence of electron pairs. A salient manifestation of flux quantization (or more precisely, fluxoid quantization) is the Little-Parks effect \cite{Little-1962} wherein the resistance of a hollow superconducting cylinder close to its critical temperature oscillates as a function of the magnetic flux inside with a periodicity given by the flux quantum $\Phi_0 = h / 2e$. In thin mesoscopic rings where vortex-crossing processes lead to a finite resistance even in the superconducting state, analogous magnetoresistance oscillations with the same periodicity are also observable much below the critical temperature due to a periodic modulation of the vortex-crossing rate \cite{Sochnikov-2010a, Sochnikov-2010b, Berdiyorov-2012, Mills-2015}.

Recently, such magnetoresistance oscillations arising from both the conventional Little-Parks effect \cite{Yasui-2017, Li-2019, Xu-2020, Li-2020} and the rate of vortex crossings \cite{Cai-2013, Cai-2020} have been identified as a useful tool in the search for exotic spin-triplet superconductivity. Promising candidate materials include Sr$_2$RuO$_4$ \cite{Maeno-1994, Maeno-2001, Mackenzie-2003, Mackenzie-2017} and UTe$_2$ \cite{Ran-2019, Aoki-2022}. In addition to the standard quantum vortices corresponding to fluxoid quantization, these unconventional superconductors may also host half-quantum vortices around which the fluxoid is quantized to a half-integer multiple of $\Phi_0$. With such half-quantum vortices present, the magnetoresistance oscillations are then expected to develop a characteristic two-peak structure \cite{Yasui-2017, Cai-2020, Vakaryuk-2011}. Importantly, half-quantum vortices are also predicted to harbor Majorana zero modes \cite{Ivanov-2001, Alicea-2012} whose non-Abelian statistics may enable intrinsically fault-tolerant quantum computation \cite{Kitaev-2003, Nayak-2008}.

In this work, we theoretically study the magnetoresistance oscillations in thin mesoscopic rings of spin-triplet superconductors below the critical temperature. Focusing on the London limit of Ginzburg-Landau theory, we adopt the formalism in Ref.~\onlinecite{Kogan-2004} to describe the available fluxoid states and thermal vortex-crossing processes by accounting for both the usual charge supercurrent and the spin supercurrent unique to spin-triplet superconductors. At the lowest temperatures, we verify that the magnetoresistance oscillates with periodicity $\Phi_0$ and has a distinctive two-peak structure \cite{Yasui-2017, Cai-2020, Vakaryuk-2011}. More interestingly, there is an intermediate temperature range in which disorder leads to magnetoresistance oscillations with a fractional periodicity $\Phi_0 / n$, where the integer $n \geq 3$ is determined by the ratio of the spin and charge superfluid densities \cite{Chung-2007}. Since these fractional oscillations directly reflect the enlarged number of available fluxoid states, we argue that they are defining hallmarks of spin-triplet superconductors, much like the integer oscillations are for their spin-singlet counterparts.

\begin{figure}[t]
\centering
\includegraphics[width=0.99\columnwidth]{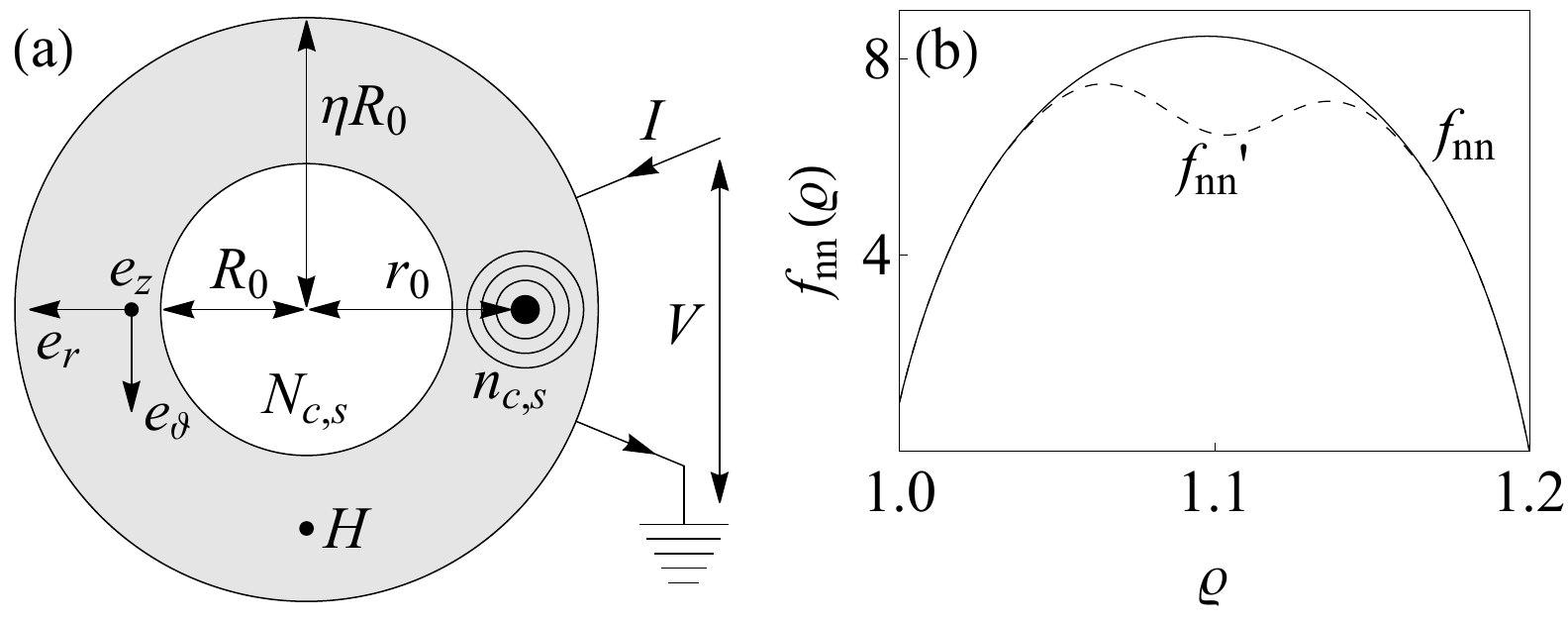}
\caption{\textbf{General setup and definitions.} (a) Thin-film superconducting ring with inner radius $R_0$ and outer radius $\eta R_0$ in a perpendicular magnetic field $\vec{H} = H \vec{e}_z$. During a snapshot of a vortex-crossing process, the central hole of the ring has charge and spin vorticities (fluxoid numbers) $N_{c,s}$, while the vortex at radius $r_0$ inside the ring has charge and spin vorticities $n_{c,s}$. Experimentally, the resistance due to such vortex-crossing processes is found by applying a bias current $I$ to a short section of the ring and measuring the voltage $V$ between the two leads. (b) Vortex self energy $f_{nn} (\varrho)$ against the vortex position $\varrho = r_0 / R_0$ for $\eta = 1.2$ without disorder (solid line) and with a single pinning site inside the ring (dashed line).}
\label{fig-1}
\end{figure}

\textbf{General formalism.} We consider a circular superconducting ring of inner radius $R_0$ and outer radius $\eta R_0$ in a perpendicular magnetic field $\vec{H} = H \vec{e}_z$ [see Fig.~\ref{fig-1}(a)]. We assume that the ring is made from a superconducting film of thickness $t \ll R_0$ and that the superconductor has spin-triplet $p_x + i p_y$ pairing with angular momentum $m_l = +1$ in real space and $m_s = \pm 1$ in spin space (with respect to the $\vec{e}_z$ direction). The spin-triplet superconducting order parameter is then \cite{Chung-2007}
\begin{equation}
\hat{\Delta} = \left[ \begin{array}{cc} \Delta_{\uparrow \uparrow} & \Delta_{\uparrow \downarrow} \\ \Delta_{\downarrow \uparrow} & \Delta_{\downarrow \downarrow} \end{array} \right] = \Delta_0 e^{i \chi} \left[ \begin{array}{cc} e^{i \alpha} & 0 \\ 0 & -e^{-i \alpha} \end{array} \right], \label{eq-Delta}
\end{equation}
where $\chi$ is the usual superconducting phase corresponding to the overall charge supercurrent, while $\alpha$ corresponds to the difference between the spin-up ($\uparrow \uparrow$) and spin-down ($\downarrow \downarrow$) supercurrents, i.e., a pure spin supercurrent. In general, the central hole of the ring has a finite vorticity (fluxoid number) for each supercurrent such that $\chi$ ($\alpha$) winds by $2\pi N_c$ ($2\pi N_s$) along the inner circumference of the ring. To understand how a vortex may travel across the ring, we further consider a vortex at position $\vec{r}_0 = (r_0, 0)$ inside the ring [see Fig.~1(a)] around which $\chi$ ($\alpha$) winds by $2\pi n_c$ ($2\pi n_s$). Importantly, the order parameter is only single valued if the two numbers within each pair $(N_c, N_s)$ and $(n_c, n_s)$ are either both integer, corresponding to a standard quantum vortex, or both half integer, corresponding to a half-quantum vortex.

Assuming $R_0 \ll \Lambda$ with the Pearl length $\Lambda = 2 \lambda^2 / t$ and the penetration depth $\lambda$, the magnetic screening inside the superconductor is negligible, and the magnetic field $\vec{B}$ is identical to the external field $\vec{H}$ \cite{Kogan-2004}. In the London limit, corresponding to a small coherence length $\xi$, the magnitude $\Delta_0$ of the order parameter at any position $\vec{r}$ further than $\xi$ from $\vec{r}_0$ is constant, and the Ginzburg-Landau free energy is then \cite{Chung-2007}
\begin{equation}
F = \frac{t \Phi_0^2} {8 \pi^2 \mu_0 \lambda^2} \int d^2 \vec{r} \left[ \big| \vec{J}_c \big|^2 + \gamma \big| \vec{J}_s \big|^2 \right] \label{eq-F}
\end{equation}
in terms of the effective charge and spin supercurrents \cite{Footnote-1}
\begin{equation}
\vec{J}_c = \vec{\nabla} \chi - \frac{2\pi} {\Phi_0} \vec{A}, \qquad \vec{J}_s = \vec{\nabla} \alpha, \label{eq-J}
\end{equation}
where the vector potential $\vec{A}$ satisfies $\vec{\nabla} \times \vec{A} = \vec{B} = \vec{H}$, while the ratio $\gamma$ of the spin and charge superfluid densities \cite{Footnote-2} is expected to be smaller than $1$ for interacting superconductors \cite{Leggett-1968, Leggett-1975} and as low as $\gamma \sim 0.3$ in a potential spin-triplet superconducting state of Sr$_2$RuO$_4$ \cite{Chung-2007}. In the absence of a bias current $I$ [see Fig.~\ref{fig-1}(a)], the charge supercurrent must satisfy the differential equations
\begin{equation}
\vec{\nabla} \cdot \vec{J}_c = 0, \quad \vec{\nabla} \times \vec{J}_c = \left[ 2\pi n_c \delta (\vec{r} - \vec{r}_0) - \frac{2 h} {R_0^2} \right] \vec{e}_z \label{eq-J-de}
\end{equation}
inside the superconductor, along with the boundary conditions
\begin{equation}
\vec{e}_n \cdot \vec{J}_c = 0, \qquad \oint_{|\vec{r}| = R_0} d\vec{r} \cdot \vec{J}_c = 2\pi \left( N_c - h \right) \label{eq-J-bc}
\end{equation}
at any interface with normal unit vector $\vec{e}_n$, and along the inner circumference of the ring, respectively, where $h = H R_0^2 \pi / \Phi_0$ is a dimensionless external field. Importantly, the spin supercurrent $\vec{J}_s$ also satisfies Eqs.~(\ref{eq-J-de}) and (\ref{eq-J-bc}) with the substitutions $n_c \to n_s$, $N_c \to N_s$, and $h \to 0$. We further note that Eqs.~(\ref{eq-J-de}) and (\ref{eq-J-bc}) are equivalent to those studied in Ref.~\onlinecite{Kogan-2004}.

Due to the linearity of Eqs.~(\ref{eq-J-de}) and (\ref{eq-J-bc}), the general solutions for the charge and spin supercurrents can be written as
\begin{equation}
\vec{J}_c = n_c \vec{J}_n + N_c \vec{J}_N - h \vec{J}_h, \quad \vec{J}_s = n_s \vec{J}_n + N_s \vec{J}_N, \label{eq-J-sol}
\end{equation}
where $\vec{J}_n$, $\vec{J}_N$, and $\vec{J}_h$ are the particular solutions of Eqs.~(\ref{eq-J-de}) and (\ref{eq-J-bc}) with $(n_c, N_c, h)$ being equal to $(1,0,0)$, $(0,1,0)$, and $(0,0,-1)$, respectively. Using polar coordinates, $\vec{r} = (r, \vartheta)$, one readily obtains $\vec{J}_N = (1 / r) \vec{e}_{\vartheta}$ and $\vec{J}_h = (r / R_0^2) \vec{e}_{\vartheta}$, while $\vec{J}_n$ for a given vortex position $r_0 = \varrho R_0$ was calculated in Ref.~\onlinecite{Kogan-2004}. Substituting Eq.~(\ref{eq-J-sol}) into Eq.~(\ref{eq-F}), the free energy of the system in the pure (vortex-free) case with $n_{c,s} = 0$ is then
\begin{equation}
F_{N_c, N_s, h}^{\mathrm{pure}} = F_0 \left[ f_{NN} \left( N_c^2 + \gamma N_s^2 \right) - 2 f_{Nh} N_c h + f_{hh} h^2 \right], \label{eq-F-pure}
\end{equation}
while in the presence of a vortex at radius $r_0 = \varrho R_0$ it reads
\begin{eqnarray}
F_{N_c, N_s, n_c, n_s, h}^{\mathrm{vortex}} (\varrho) = F_{N_c, N_s, h}^{\mathrm{pure}} + F_0 \big[ f_{nn} (\varrho) \left( n_c^2 + \gamma n_s^2 \right) && \nonumber \\
+ 2 f_{nN} (\varrho) \left( n_c N_c + \gamma n_s N_s \right) - 2 f_{nh} (\varrho) \, n_c h \big], \quad && \label{eq-F-vortex}
\end{eqnarray}
where $F_0 = t \Phi_0^2 \ln \eta / (4 \pi \mu_0 \lambda^2)$ is an overall energy scale, and $f_{XY} = (2\pi \ln \eta)^{-1} \int d^2 \vec{r} \, \vec{J}_X \cdot \vec{J}_Y$ ($X,Y = n,N,h$) are dimensionless free energies \cite{Kogan-2004}:
\begin{eqnarray}
f_{NN} = 1, \quad f_{Nh} = \frac{\eta^2 - 1} {2 \ln \eta}, \quad f_{hh} = \frac{\eta^4 - 1} {4 \ln \eta}, && \nonumber \\
f_{nN} (\varrho) = 1 - \frac{\ln \varrho} {\ln \eta}, \quad f_{nh} (\varrho) = \frac{\eta^2 - \varrho^2} {2 \ln \eta}, \, && \label{eq-f}
\end{eqnarray}
while $f_{nn} (\varrho)$ has the form plotted in Fig.~\ref{fig-1}(b). We remark that $f_{nn} (\varrho)$, corresponding to the self energy of the vortex, nominally diverges in the London limit and must be regularized with a small but finite coherence length $\xi$ \cite{Footnote-3}.

\begin{figure*}[t]
\centering
\includegraphics[width=1.98\columnwidth]{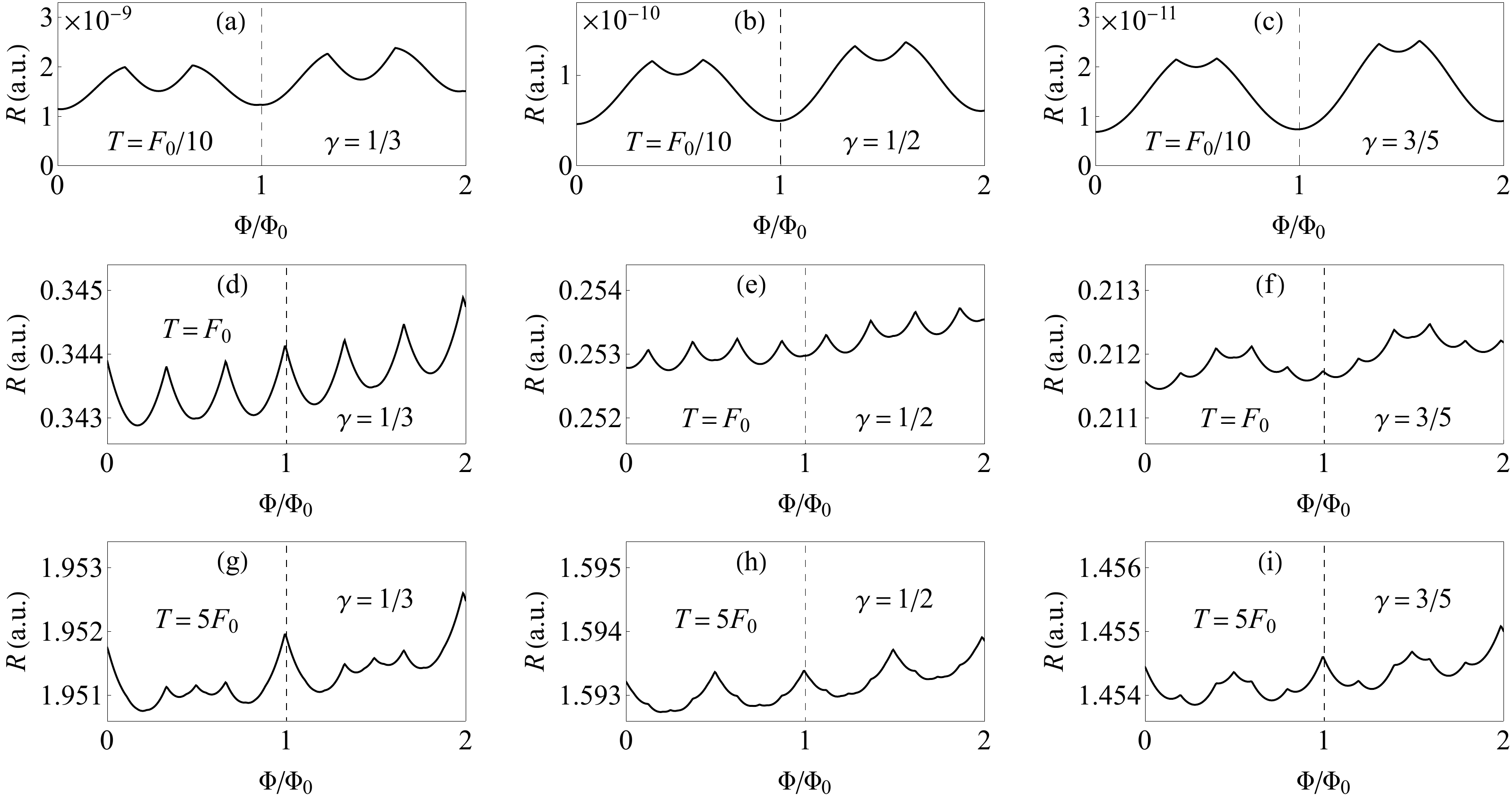}
\caption{\textbf{Magnetoresistance oscillations at different temperatures.} Resistance $R$ of the superconducting ring in Fig.~\ref{fig-1}(a), as calculated from Eq.~(\ref{eq-R}), against the dimensionless flux $\phi = \Phi / \Phi_0$ at low temperatures $T = F_0 / 10$ (a-c), intermediate temperatures $T = F_0$ (d-f), and high temperatures $T = 5 F_0$ (g-i) [in terms of $F_0 = t \Phi_0^2 \ln \eta / (4 \pi \mu_0 \lambda^2)$] for a radius ratio $\eta = 1.2$ and superfluid-density ratios $\gamma = 1/3$ (a,d,g), $\gamma = 1/2$ (b,e,h), and $\gamma = 3/5$ (c,f,i) in the presence of a single pinning site inside the ring [see Fig.~\ref{fig-1}(b)].}
\label{fig-2}
\end{figure*}

\textbf{Theory of magnetoresistance.} We first assume that the superconducting ring in Fig.~\ref{fig-1}(a) is in thermal equilibrium without any bias current $I$. Because of the large vortex self energy in the London limit, there are no stable vortices inside the superconductor at sufficiently low temperatures. Nonetheless, at any finite temperature $T = 1 / \beta$, the fluxoid numbers $N_{c,s}$ of the central hole can thermally fluctuate, and the probability of the system to be in the fluxoid state $(N_c, N_s)$ is given by
\begin{equation}
P_{(N_c, N_s)} = \frac{1}{Z} \exp \left[ -\beta F_{N_c, N_s, h}^{\mathrm{pure}} \right], \label{eq-P}
\end{equation}
where $Z = \sum_{N_c, N_s} \exp [-\beta F_{N_c, N_s, h}^{\mathrm{pure}}]$. The thermal fluctuations themselves happen by vortices traveling across the ring; the fluxoid state of the system transitions from $(N_c, N_s)$ to $(N_c', N_s')$ if a vortex with $(n_c, n_s) = \kappa (N_c'-N_c, N_s'-N_s)$ and $\kappa = +1$ ($\kappa = -1$) crosses the ring in the inward (outward) direction. If these two processes are thermally activated, their respective free-energy barriers are \cite{Kogan-2004}
\begin{eqnarray}
F_{(N_c, N_s) \to (N_c', N_s'), h}^{\mathrm{barrier}, +} &=& \max_{\varrho} F_{N_c, N_s, N_c'-N_c, N_s'-N_s, h}^{\mathrm{vortex}} (\varrho) \nonumber \\
&& - F_{N_c, N_s, h}^{\mathrm{pure}}, \label{eq-F-barrier} \\
F_{(N_c, N_s) \to (N_c', N_s'), h}^{\mathrm{barrier}, -} &=& \max_{\varrho} F_{N_c', N_s', N_c-N_c', N_s-N_s', h}^{\mathrm{vortex}} (\varrho) \nonumber \\
&& - F_{N_c, N_s, h}^{\mathrm{pure}}, \nonumber
\end{eqnarray}
and the total transition rate from $(N_c, N_s)$ to $(N_c', N_s')$ is then
\begin{eqnarray}
\Gamma_{(N_c, N_s) \to (N_c', N_s'), h} &=& P_{(N_c, N_s)} \, A_{(N_c, N_s) \to (N_c', N_s'), h}, \label{eq-Gamma} \\
A_{(N_c, N_s) \to (N_c', N_s'), h} &\propto& \sum_{\pm} \exp \left[ -\beta F_{(N_c, N_s) \to (N_c', N_s'), h}^{\mathrm{barrier}, \pm} \right]. \nonumber
\end{eqnarray}
We note that, in thermal equilibrium, detailed balance is satisfied: $\Gamma_{(N_c, N_s) \to (N_c', N_s'), h} = \Gamma_{(N_c', N_s') \to (N_c, N_s), h}$.

\begin{figure*}[t]
\centering
\includegraphics[width=1.98\columnwidth]{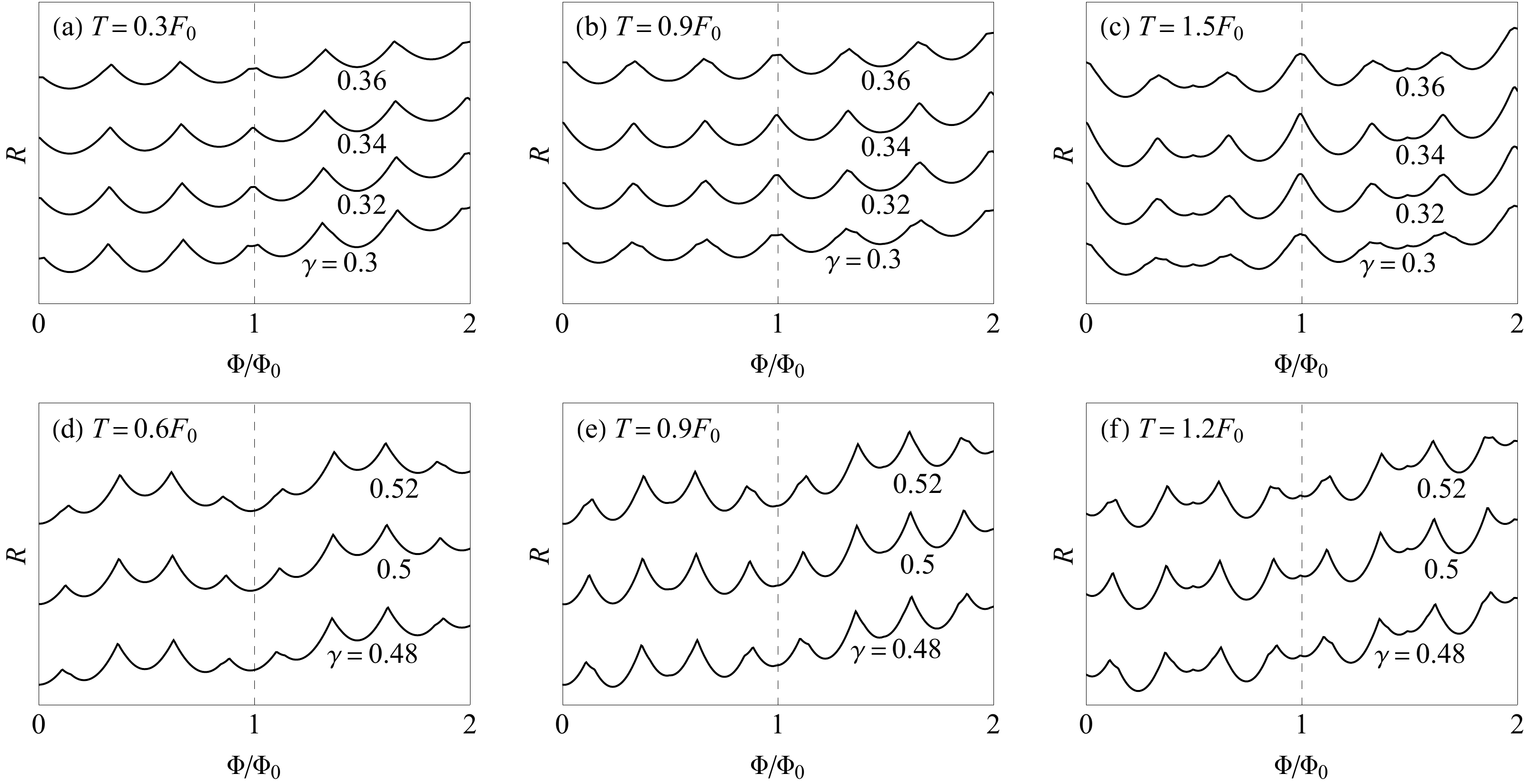}
\caption{\textbf{Robustness of fractional oscillations.} Magnetoresistance oscillations with fractional periodicities $\Delta \phi = 1/3$ (a-c) and $\Delta \phi = 1/4$ (d-f) in the intermediate temperature ranges $0.3 \leq T / F_0 \leq 1.5$ and $0.6 \leq T / F_0 \leq 1.2$ for superfluid-density ratios $0.3 \leq \gamma \leq 0.36$ and $0.48 \leq \gamma \leq 0.52$, respectively. In each case, the resistance $R$ of the superconducting ring in Fig.~\ref{fig-1}(a) is calculated from Eq.~(\ref{eq-R}) against the dimensionless flux $\phi = \Phi / \Phi_0$ for a radius ratio $\eta = 1.2$ in the presence of a single pinning site inside the ring [see Fig.~\ref{fig-1}(b)]. The different curves are labeled by $\gamma$ and are vertically shifted with respect to each other for better visibility.}
\label{fig-3}
\end{figure*}

Next, we assume that a bias current $I$ is applied by attaching two leads to the superconducting ring [see Fig.~\ref{fig-1}(a)]. For each vortex with a given sign of the charge vorticity $n_c$, the bias current exerts a force in the inward or outward direction, thus leading to a net flow of such vortices in one of these directions by decreasing the free-energy barrier in one direction and increasing it in the other one. The resulting rate of phase slips then gives rise to a finite voltage between the two leads and translates into a finite resistance for the superconducting ring \cite{Halperin-2010}. Without affecting our main results, we make a simplifying assumption that the two leads are close to each other along the ring [see Fig.~\ref{fig-1}(a)]. In this case, the entire bias current goes through the short section of the ring between the two leads, and the probabilities $P_{(N_c, N_s)}$ of the fluxoid states are still given by Eq.~(\ref{eq-P}). However, from the perspective of the transition rates $A_{(N_c, N_s) \to (N_c', N_s'), h}$ within the short section, the charge fluxoid number is effectively reduced by $\varepsilon = I / I_0$, where $I_0 = t \Phi_0 \ln \eta / (2\pi \mu_0 \lambda^2)$. Hence, for a small bias current $I \ll I_0$, the resistance between the two leads becomes
\begin{equation}
R \propto \sum_{N_c, N_s} P_{(N_c, N_s)} \sum_{n_c, n_s} n_c \, \frac{\partial A_{(N_c - \varepsilon, N_s) \to (\tilde{N}_c - \varepsilon, \tilde{N}_s), h}} {\partial \varepsilon} \Bigg|_{\varepsilon = 0}, \label{eq-R}
\end{equation}
where $\tilde{N}_{c,s} \equiv N_{c,s} + n_{c,s}$, while $A_{(N_c - \varepsilon, N_s) \to (\tilde{N}_c - \varepsilon, \tilde{N}_s), h}$ for $\varepsilon \neq 0$ is computed through Eqs.~(\ref{eq-F-barrier}) and (\ref{eq-Gamma}) by formally evaluating Eqs.~(\ref{eq-F-pure}) and (\ref{eq-F-vortex}) at a fractional value of $N_c$. Finally, to obtain our full set of main results, we assume that the short section of the ring between the two leads contains some form of disorder. For concreteness, we consider a single localized ``pinning site'' (e.g., defect or impurity) that renormalizes the vortex self energy from $f_{nn} (\varrho)$ to $f_{nn}' (\varrho)$ [see Fig.~\ref{fig-1}(b)].

\textbf{Results and discussion.} The resistance $R$ of the superconducting ring is plotted in Fig.~\ref{fig-2} against the external field $H$ for different values of the temperature $T$ and the superfluid-density ratio $\gamma$. We parameterize the external field in terms of the dimensionless flux $\phi = \Phi / \Phi_0$, where $\Phi = H R_{\mathrm{eff}}^2 \pi$ is the flux inside the effective mean radius \cite{Kogan-2004}
\begin{equation}
R_{\mathrm{eff}} = R_0 \sqrt{\frac{f_{Nh}} {f_{NN}}} = R_0 \sqrt{\frac{\eta^2 - 1} {2 \ln \eta}}. \label{eq-R-eff}
\end{equation}
In this parameterization, conventional magnetoresistance oscillations in spin-singlet superconductors \cite{Sochnikov-2010a, Sochnikov-2010b, Berdiyorov-2012, Mills-2015} have unit periodicity $\Delta \phi = 1$ with a peak at each external field $\phi = N + 1/2$ ($N \in \mathbb{Z}$). In contrast, Fig.~\ref{fig-2} shows that spin-triplet superconductors with $\gamma < 1$ possess nontrivial additional structure in their magnetoresistance oscillations. For the lowest temperatures ($T \ll F_0$), the periodicity is still $\Delta \phi = 1$, but each peak at $\phi = N + 1/2$ splits into two peaks that move further apart as $\gamma$ is decreased \cite{Yasui-2017, Cai-2020, Vakaryuk-2011}. For high temperatures ($T \gg F_0$), the oscillations are significantly more complex with an overall periodicity $\Delta \phi = 1$ or $\Delta \phi = 1/2$. Most interestingly, for intermediate temperatures ($T \sim F_0$), the magnetoresistance oscillations have an emergent fractional periodicity $\Delta \phi = 1/n$, where the integer $n$ is determined by the superfluid-density ratio $\gamma$. While Fig.~\ref{fig-2} suggests that the different integers $n \geq 3$ correspond to specific rational values of $\gamma$, it is also demonstrated in Fig.~\ref{fig-3} that the fractional periodicities $\Delta \phi = 1/n$ persist in finite ranges of both $\gamma$ and $T$.

\begin{figure*}[t]
\centering
\includegraphics[width=1.98\columnwidth]{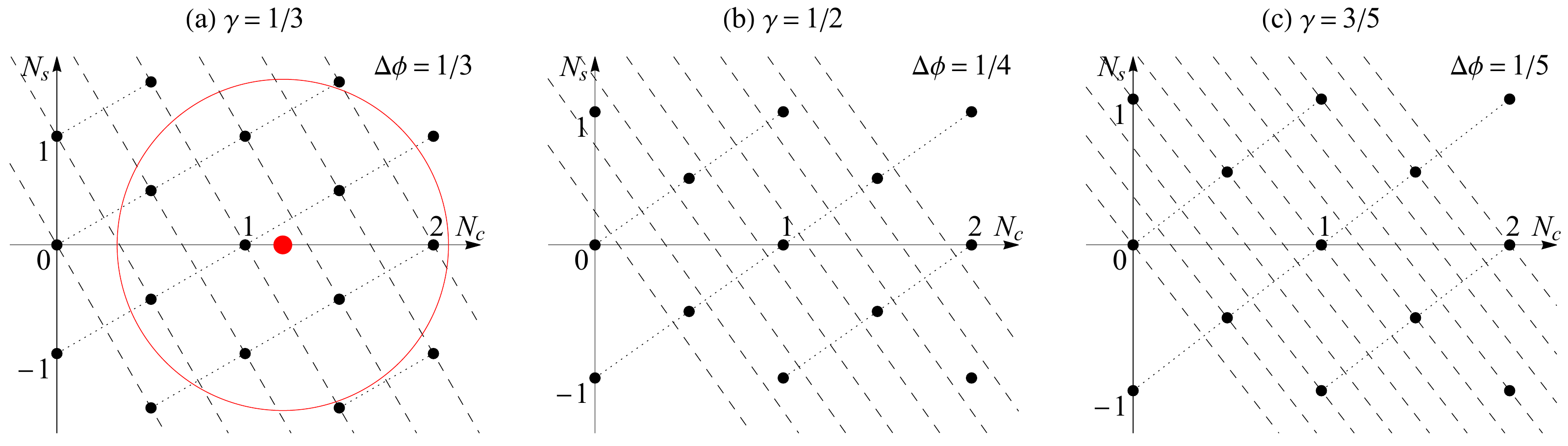}
\caption{\textbf{Geometric interpretation of fractional oscillations.} Emergence of the fractional periodicities $\Delta \phi = 1/3$ (a),  $\Delta \phi = 1/4$ (b), and $\Delta \phi = 1/5$ (c) from the superfluid-density ratios $\gamma = 1/3$, $\gamma = 1/2$, and $\gamma = 3/5$, respectively. Within a two-dimensional plane, the black dots depict the possible fluxoid states $(N_c, N_s)$, while the red dot at position $(\phi, 0)$ represents the external field. Due to the scaling factor $\sqrt{\gamma}$ between the vertical ($N_s$) and horizontal ($N_c$) dimensions, the energy of a given fluxoid state is proportional to the distance squared between the corresponding black dot and the red dot [see Eq.~(\ref{eq-F-pure-new})]. Focusing on the half-quantum transitions $n_{c,s} = \Delta N_{c,s} = 1/2$ (dotted lines), the argument $(N_c + \gamma N_s - \phi) / 2$ of $G_{1/2, 1/2}$ in Eq.~(\ref{eq-R-new}) corresponds to the perpendicular projection of the red dot onto the dotted line connecting $(N_c, N_s)$ and $(N_c + 1/2, N_s + 1/2)$. Therefore, as the external field $\phi$ is increased, the same feature in the magnetoresistance is periodically replicated every time the red dot at position $(\phi, 0)$ crosses a perpendicular bisector (dashed line). Relevant transitions connecting fluxoid states with sizeable probabilities are within the red circle whose radius scales with the square root of the temperature.}
\label{fig-4}
\end{figure*}

To understand these fractional oscillations, we first notice that the free energy of a pure (vortex-free) system in Eq.~(\ref{eq-F-pure}) can be written in the new parameterization as
\begin{equation}
F_{N_c, N_s, \phi}^{\mathrm{pure}} = F_0 \left[ \left( N_c - \phi \right)^2 + \gamma N_s^2 + g (\phi) \right]. \label{eq-F-pure-new}
\end{equation}
For external field $\phi$, the free-energy difference between two fluxoid states $(N_c, N_s)$ and $(\tilde{N}_c, \tilde{N}_s) = (N_c + n_c, N_s + n_s)$, connected by vortices $\pm (n_c, n_s)$ crossing the ring, is then
\begin{eqnarray}
F_{\tilde{N}_c, \tilde{N}_s, \phi}^{\mathrm{pure}} - F_{N_c, N_s, \phi}^{\mathrm{pure}} = 2 F_0 \left[ n_c \left( N_c - \phi \right) + \gamma n_s N_s \right] && \nonumber \\
+ F_0 \left( n_c^2 + \gamma n_s^2 \right). \qquad \qquad \,\, && \label{eq-F-diff}
\end{eqnarray}
Moreover, if the radius ratio $\eta$ of the superconducting ring is not too large, $f_{nN} (\varrho)$ and $f_{nh} (\varrho)$ in Eq.~(\ref{eq-f}) are close to linear for $1 \leq \varrho \leq \eta$. Hence, taking a linear interpolation between their values at $\varrho = 1$ and $\varrho = \eta$, the free energy of the system with a single vortex [see Eq.~(\ref{eq-F-vortex})] can be approximated by
\begin{eqnarray}
F_{N_c, N_s, n_c, n_s, \phi}^{\mathrm{vortex}} (\varrho) = F_{N_c, N_s, \phi}^{\mathrm{pure}} + F_0 f_{nn}' (\varrho) \left( n_c^2 + \gamma n_s^2 \right) \qquad && \label{eq-F-vortex-new} \\
\qquad \qquad + 2 F_0 \left[ n_c \left( N_c - \phi \right) + \gamma n_s N_s \right] \frac{\eta - \varrho} {\eta - 1}. \quad && \nonumber
\end{eqnarray}
Importantly, if we use this approximation, the transition rates $A_{(N_c, N_s) \to (N_c + n_c, N_s + n_s), \phi}$ in Eq.~(\ref{eq-Gamma}) only depend on either $\phi$ or $N_{c,s}$ via the combination $n_c (N_c - \phi) + \gamma n_s N_s$, and the resistance in Eq.~(\ref{eq-R}) thus takes the general form
\begin{equation}
R \propto \sum_{N_c, N_s} P_{(N_c, N_s)} \sum_{n_c, n_s} G_{n_c, n_s} \left[ n_c \left( N_c - \phi \right) + \gamma n_s N_s \right]. \label{eq-R-new}
\end{equation}
Due to the many identical contributions $G_{n_c, n_s}$ corresponding to different $N_{c,s}$, each shifted by $N_c + \gamma N_s n_s / n_c$ in the field $\phi$, this form naturally leads to periodic oscillations.

Next, we recall from Eq.~(\ref{eq-F-vortex-new}) that the vortex self energy is proportional to $n_c^2 + \gamma n_s^2$. For any $\gamma < 1$, the dominant vortices contributing to the resistance \cite{Footnote-4} at sufficiently low temperatures [$T \ll F_0 \max_{\varrho} f_{nn}' (\varrho)$] are then the half-quantum vortices with $n_{c,s} = \pm 1/2$. In the intermediate temperature range ($T \sim F_0$), there are also many fluxoid states $(N_c, N_s)$ with sizeable probabilities $P_{(N_c, N_s)} \sim 1$. If we then sum over the identical contributions $G_{\pm 1/2, \pm 1/2}$ in Eq.~(\ref{eq-R-new}) for \emph{all possible} $N_{c,s}$, each shifted by $N_c \pm \gamma N_s$ in the field $\phi$, these identical contributions conspire to produce fractional oscillations with periodicity $\Delta \phi = 1/n$. For a rational value of the superfluid-density ratio, $\gamma = p/q$, with the integers $p$ and $q$ being relative primes, it is shown in the Supplemental Material (SM) \cite{SM} that $n = q$ if $p$ and $q$ are both odd and $n = 2q$ otherwise. Thus, in accordance with Fig.~\ref{fig-2}, the fractional periodicities are $\Delta \phi = 1/3$, $\Delta \phi = 1/4$, and $\Delta \phi = 1/5$ for $\gamma = 1/3$, $\gamma = 1/2$, and $\gamma = 3/5$, respectively. In practice, since the summation over $N_{c,s}$ is cut off at any finite temperature $T \sim F_0$, only the fractional periodicities with small $p$ and $q$ are observable, but each of them remains observable in a finite range around $\gamma = p/q$ (see Fig.~\ref{fig-3}). We further remark that the emergence of fractional oscillations and the intimate connection between $\Delta \phi$ and $\gamma$ can also be understood from a simple geometric picture (see Fig.~\ref{fig-4}).

Interestingly, as it is demonstrated in the SM \cite{SM}, the fractional magnetoresistance oscillations only appear if disorder is present in the superconductor. While the fractional periodicity $\Delta \phi = 1/n$ itself is a robust emergent feature connected to the superfluid-density ratio $\gamma$, the corresponding oscillations are not observable if the functions $G_{\pm 1/2, \pm 1/2}$ are completely smooth. The crucial role of disorder is to produce nonanalytic features in $G_{\pm 1/2, \pm 1/2}$ that can be replicated periodically as a function of the field $\phi$. For the specific form of disorder considered in this work (i.e., a single pinning site), it is illustrated in the SM \cite{SM} how a discontinuity in the vortex position $\varrho_0$ corresponding to the maximum of the vortex energy function $F_{N_c, N_s, 1/2, 1/2, \phi}^{\mathrm{vortex}} (\varrho)$ leads to a cusp in $G_{1/2, 1/2}$.

We finally remark that, as the temperature $T$ approaches the critical temperature of the superconductor, the effective temperature $T / F_0$ with $F_0 = t \Phi_0^2 \ln \eta / (4 \pi \mu_0 \lambda^2)$ diverges as a result of $\lambda \to \infty$. Therefore, in principle, the intermediate temperatures $T \sim F_0$ that give rise to the fractional magnetoresistance oscillations are attainable for any ring dimensions. In practice, however, we expect the fractional oscillations to be more observable further away from the critical temperature, which is achieved by keeping both the film thickness $t$ and the radius ratio $\eta$ as small as possible.

\vspace{0.4cm}

\noindent{\textbf{Data availability}}

\noindent{The data that support the findings of this study are available from the author upon reasonable request.}

\vspace{0.3cm}

\noindent{\textbf{Code availability}}

\noindent{The codes that support the findings of this study are available from the author upon reasonable request.}

\vspace{0.3cm}

\noindent{\textbf{Acknowledgements}}

\noindent{We thank Benjamin Lawrie and Yun-Yi Pai for experimental motivation as well as Eugene Dumitrescu and Chengyun Hua for helpful discussions. This research was sponsored by the U.~S.~Department of Energy, Office of Science, Basic Energy Sciences, Materials Sciences and Engineering Division.}

\vspace{0.3cm}

\noindent{\textbf{Competing interests}}

\noindent{The author declares no competing interests.}

\clearpage

\begin{widetext}

\begin{center}
{\large \textbf{Supplemental Material}}
\end{center}

\section{Derivation of the fractional periodicity}

Here we aim to determine the fractional periodicity of the magnetoresistance oscillations that emerges when summing over the identical contributions $G_{1/2, 1/2} [\{(N_c - \phi) + \gamma N_s\} / 2]$ for all $N_c$ and $N_s$ in Eq.~(18) of the main text. We note that the same periodicity is also obtained when summing over the symmetry-related contributions $G_{1/2, -1/2}$, $G_{-1/2, 1/2}$, and $G_{-1/2, -1/2}$. By recognizing that each contribution $G_{1/2, 1/2} [\{(N_c - \phi) + \gamma N_s\} / 2]$ has a relative shift $\delta \phi_{N_c, N_s} = N_c + \gamma N_s$ in the external field $\phi$, the periodicity $\Delta \phi$ can be determined by finding the set of all possible shifts $\{\delta \phi_{N_c, N_s}\}$ and taking the smallest possible difference between any two shifts within this set. We assume that the superfluid-density ratio $\gamma$ is a rational number and can be written as $\gamma = p/q$ with the integers $p$ and $q$ being relative primes.

The summation over $N_c$ and $N_s$ is complicated by the fact that these two variables are not independent from each other: they are either both integer or both half integer. Thus, it is useful to express them as $N_c = (N_{\uparrow} + N_{\downarrow}) / 2$ and $N_s = (N_{\uparrow} - N_{\downarrow}) / 2$ in terms of the independent variables $N_{\uparrow}$ and $N_{\downarrow}$ that can both take arbitrary integer values. The set of all shifts is then given by
\begin{equation}
\left\{ \delta \phi_{N_c, N_s} \right\} = \left\{ \frac{(q + p) N_{\uparrow} + (q - p) N_{\downarrow}} {2q} \, \Bigg| \, N_{\uparrow, \downarrow} \in \mathbb{Z} \right\}. \label{eq-phi}
\end{equation}
In the following, we consider two cases based on the parities of the integers $p$ and $q$. Recognizing that they cannot both be even (as they are assumed to be relative primes), they are either both odd or one of them is even and the other one is odd.

\emph{First case: $p$ and $q$ are both odd.} Since $q + p$ and $q - p$ are both even in this case, $r = (q + p) / 2$ and $s = (q - p) / 2$ are both integers. Furthermore, $r$ and $s$ must be relative primes. Indeed, if we assume that they are not relative primes, they have a common prime factor $z$ and can be written as $r = z r'$ and $s = z s'$ with $r', s' \in \mathbb{Z}$. Then, we can write $p = r - s = z (r' - s')$ and $q = r + s = z (r' + s')$, which contradicts our initial assumption that $p$ and $q$ are relative primes.

In terms of $r$ and $s$, the set of all shifts in Eq.~(\ref{eq-phi}) becomes $\{\delta \phi_{N_c, N_s}\} = \{ (r N_{\uparrow} + s N_{\downarrow}) / q \, | \, N_{\uparrow, \downarrow} \in \mathbb{Z} \}$. We first notice that all numbers in this set must be of the form $m / q$ with $m \in \mathbb{Z}$. We can also show that all numbers of this form are in the set $\{\delta \phi_{N_c, N_s}\}$ by proving that, for any integer $m$ and any pair of relative prime integers $r$ and $s$, there exist integers $N_{\uparrow}$ and $N_{\downarrow}$ such that $m = r N_{\uparrow} + s N_{\downarrow}$. We start the proof by considering the $s$-element set $\{ 0, r, 2r, \ldots, (s-1) r\}$ in which all elements must have different moduli with respect to $s$. Indeed, if we assume that two elements $a r$ and $b r > a r$ with $a, b \in \mathbb{Z}$ have the same modulus with respect to $s$, the number $(b - a) r < s r$ is divisible by $s$, which contradicts the initial assumption that $r$ and $s$ are relative primes. In turn, if the $s$ elements all have different moduli with respect to $s$, they realize all possible moduli with respect to $s$, and one of them must have the same modulus as $m$ with respect to $s$. If we then call this element $N_{\uparrow} r$ and define $N_{\downarrow} = (m - N_{\uparrow} r) / s$, the integers $N_{\uparrow}$ and $N_{\downarrow}$ clearly satisfy $m = r N_{\uparrow} + s N_{\downarrow}$, which concludes our proof.

Finally, since the set of all shifts can be written as $\{\delta \phi_{N_c, N_s}\} = \{ m / q \, | \, m \in \mathbb{Z} \}$, the emergent periodicity of the magnetoresistance oscillations is readily found to be $\Delta \phi = 1 / q$ as claimed in the main text.

\emph{Second case: $p$ and $q$ have opposite parities.} In this case, the integers $\tilde{r} = q + p$ and $\tilde{s} = q - p$ are both odd. Furthermore, $\tilde{r}$ and $\tilde{s}$ must be relative primes. Indeed, if we assume that they are not relative primes, they have a common prime factor $w > 2$ and can be written as $\tilde{r} = w \tilde{r}'$ and $\tilde{s} = w \tilde{s}'$, where $\tilde{r}'$ and $\tilde{s}'$ are odd integers. Then, by defining the integers $\tilde{p} = (\tilde{r}' - \tilde{s}') / 2$ and $\tilde{q} = (\tilde{r}' + \tilde{s}') / 2$, we can write $p = (\tilde{r} - \tilde{s}) / 2 = w \tilde{p}$ and $q = (\tilde{r} + \tilde{s}) / 2 = w \tilde{q}$, which contradicts our initial assumption that $p$ and $q$ are relative primes.

In terms of $\tilde{r}$ and $\tilde{s}$, the set of all shifts in Eq.~(\ref{eq-phi}) is given by $\{\delta \phi_{N_c, N_s}\} = \{ (\tilde{r} N_{\uparrow} + \tilde{s} N_{\downarrow}) / (2q) \, | \, N_{\uparrow, \downarrow} \in \mathbb{Z} \}$. We first notice that all numbers in this set must be of the form $m / (2q)$ with $m \in \mathbb{Z}$. We can also show that all numbers of this form are in the set $\{\delta \phi_{N_c, N_s}\}$ by invoking the same proof as in the first case above.

Finally, since the set of all shifts can be written as $\{\delta \phi_{N_c, N_s}\} = \{ m / (2q) \, | \, m \in \mathbb{Z} \}$, the emergent periodicity of the magnetoresistance oscillations is readily found to be $\Delta \phi = 1 / (2q)$ as claimed in the main text.

\begin{figure}[t]
\centering
\includegraphics[width=0.99\columnwidth]{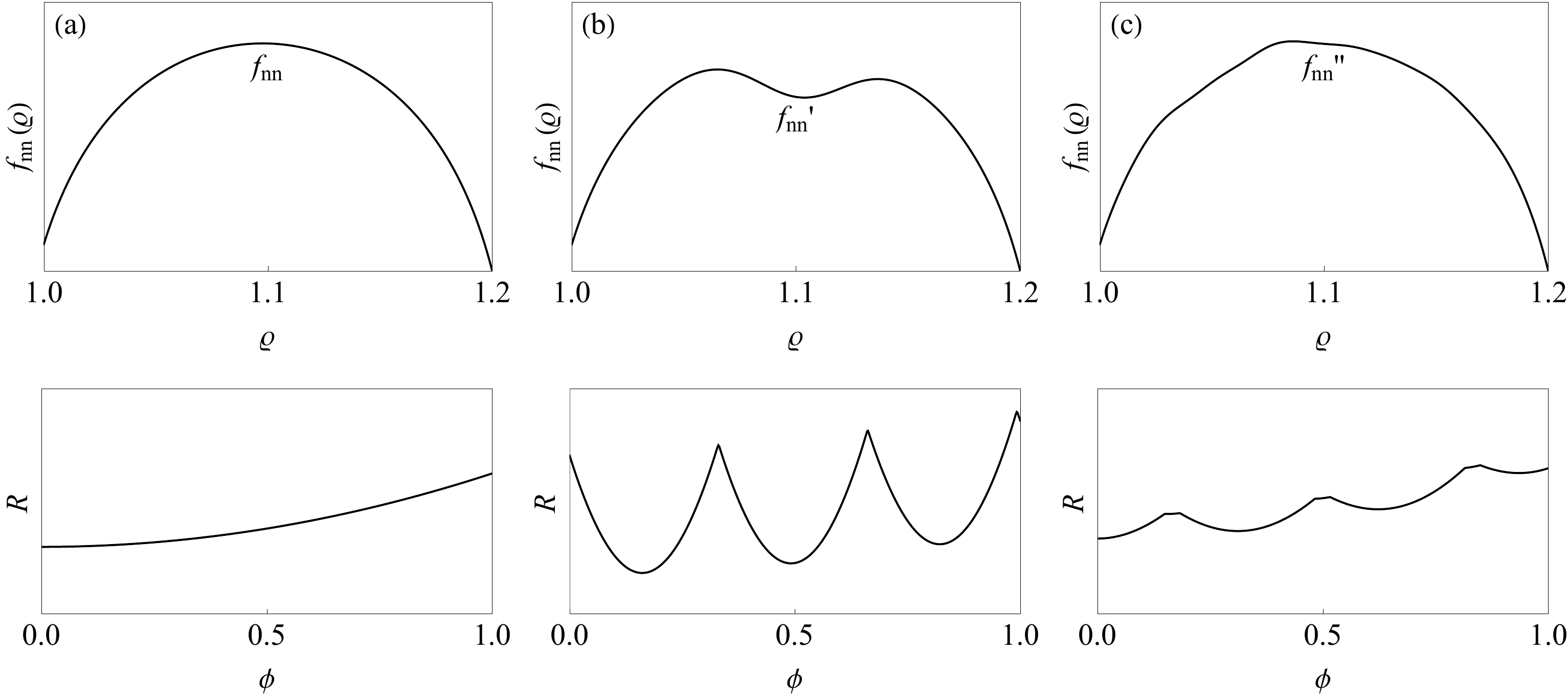}
\caption{Vortex self energy $f_{nn} (\varrho)$ against the dimensionless vortex position $\varrho$ (top) and the corresponding resistance $R$ of the superconducting ring against the external field (i.e., dimensionless flux) $\phi = \Phi / \Phi_0$ at the intermediate temperature $T = F_0 / 2$ (bottom) for radius ratio $\eta = 1.2$ and superfluid-density ratio $\gamma = 1/3$ without any disorder (a), with a single pinning site (b), and with a random potential landscape (c).}
\label{fig-5}
\end{figure}

\section{Connection between disorder and fractional oscillations}

Here we demonstrate that the fractional magnetoresistance oscillations only appear in the presence of disorder and illustrate a specific mechanism by which disorder can give rise to periodically replicated peaks in the magnetoresistance. In Fig.~\ref{fig-5}, the magnetoresistance is plotted without any disorder, with the same kind of disorder as in the main text (single pinning site), and with a completely different kind of disorder (random potential landscape). While the magnetoresistance is featureless in the first case, it exhibits fractional oscillations with the same periodicity in the remaining two cases.

In the following, we restrict our attention to the second case and assume that the only source of disorder is a single pinning site inside the ring that renormalizes the vortex self energy from $f_{nn} (\varrho)$ to $f_{nn}' (\varrho)$ [see Fig.~\ref{fig-5}(b)]. We describe how this kind of disorder gives nonanalytic features (cusps) in the transition rate $A_{(0, 0) \to (1/2, 1/2), \phi}$ [see Eq.~(12) in the main text] and hence the function $G_{1/2, 1/2}$ [see Eq.~(18) of the main text] that manifest as peaks in the magnetoresistance and are periodically replicated as a function of the external field $\phi$ to produce fractional magnetoresistance oscillations.

\begin{figure}[b]
\centering
\includegraphics[width=0.99\columnwidth]{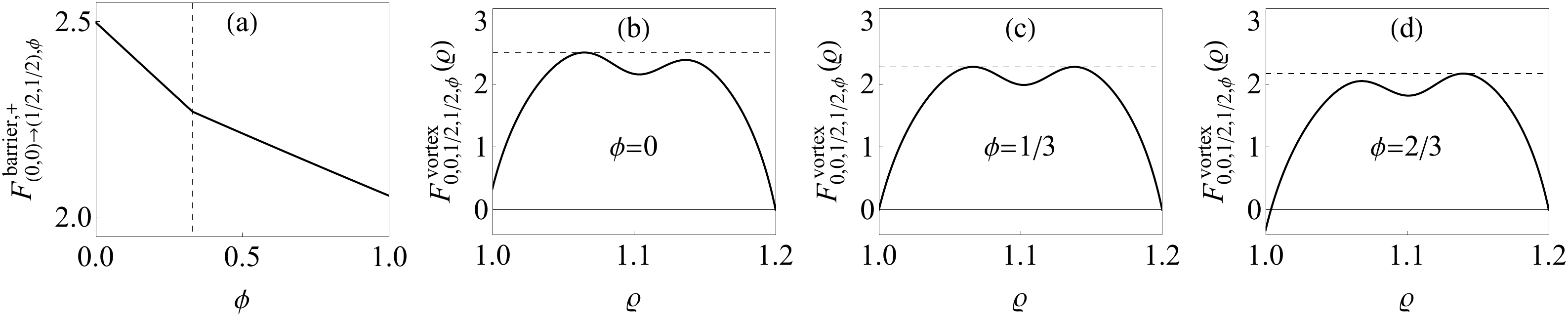}
\caption{(a) Vortex energy barrier $F_{(0, 0) \to (1/2, 1/2), \phi}^{\mathrm{barrier}, +}$ against the external field (i.e., dimensionless flux) $\phi = \Phi / \Phi_0$ for radius ratio $\eta = 1.2$ and superfluid-density ratio $\gamma = 1/3$ in the presence of a single pinning site inside the ring [see Fig.~\ref{fig-5}(b)]. The dashed line indicates the critical field $\phi_0^{+} \approx 1/3$ at which the first derivative has a discontinuity. (b-d) Vortex energy function $F_{0, 0, 1/2, 1/2, \phi}^{\mathrm{vortex}} (\varrho)$ against the dimensionless vortex position $\varrho$ for three different external fields: $\phi = 0$ (b), $\phi = 1/3$ (c), and $\phi = 2/3$ (d). In each case, the dashed line marks the maximum of the vortex energy function, i.e., the vortex energy barrier shown in subfigure (a). The critical field $\phi_0^{+} \approx 1/3$ corresponds to a discontinuity in the vortex position $\varrho_0^{+}$ that maximizes the vortex energy function.}
\label{fig-6}
\end{figure}

\begin{figure}[t]
\centering
\includegraphics[width=0.95\columnwidth]{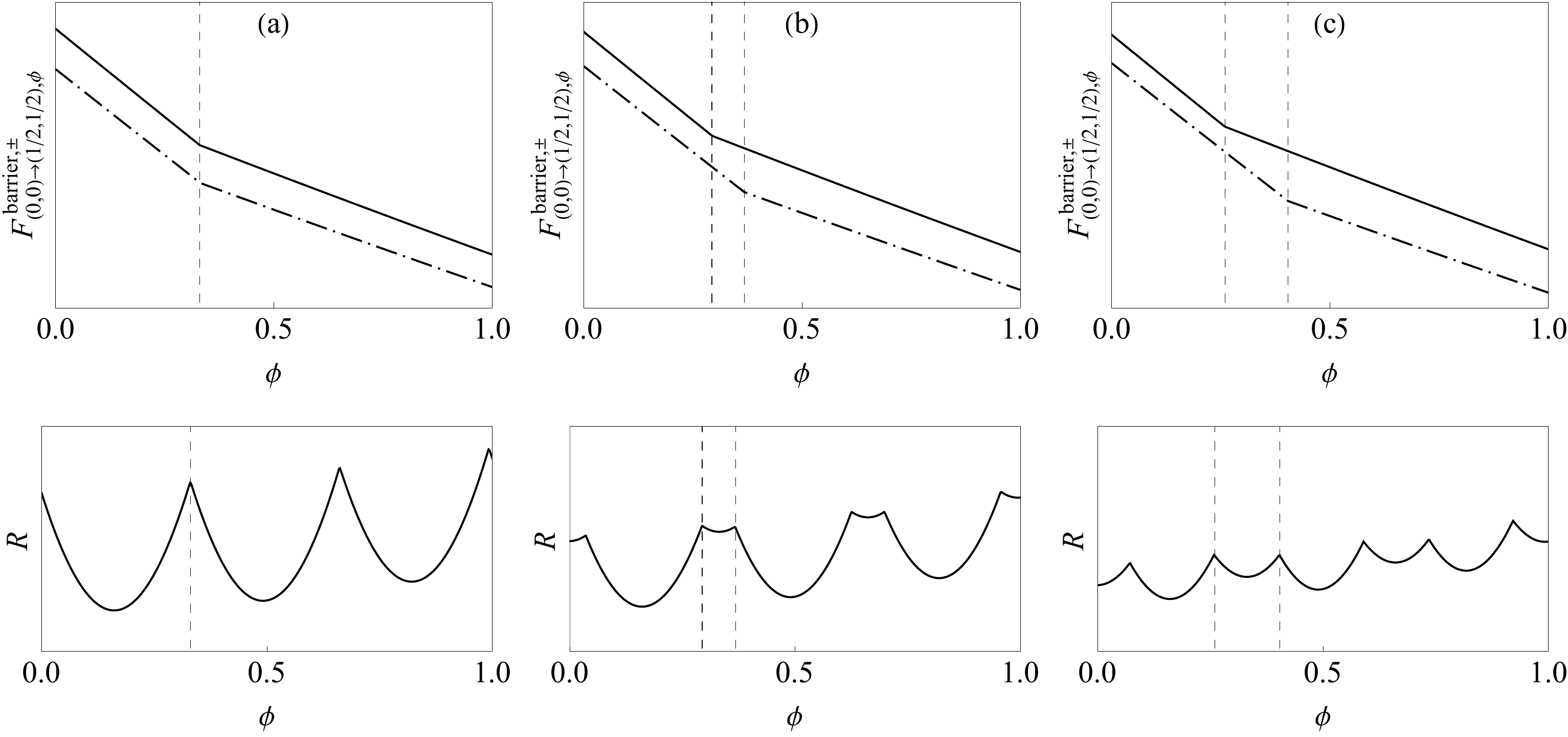}
\caption{Vortex energy barriers $F_{(0, 0) \to (1/2, 1/2), \phi}^{\mathrm{barrier}, \pm}$ (top) and the corresponding resistance $R$ of the superconducting ring at the intermediate temperature $T = F_0 / 2$ (bottom) against the external field (i.e., dimensionless flux) $\phi = \Phi / \Phi_0$ for radius ratio $\eta = 1.2$ and superfluid-density ratio $\gamma = 1/3$ in the presence of a single pinning site at (a) the same location as in Fig.~\ref{fig-5}(b) and (b-c) progressively moved in the inward direction. The two vortex energy barriers $F_{(0, 0) \to (1/2, 1/2), \phi}^{\mathrm{barrier}, +}$ (solid line) and $F_{(0, 0) \to (1/2, 1/2), \phi}^{\mathrm{barrier}, -}$ (dash-dotted line) in the top panels are vertically shifted with respect to each other for better visibility. In each case, the dashed lines indicate the two critical fields $\phi_0^{\pm}$ that correspond to cusps in the vortex energy barriers and peaks replicated with periodicity $\Delta \phi = 1/3$ in the magnetoresistance.}
\label{fig-7}
\end{figure}

From Eq.~(12) in the main text, the transition rate $A_{(0, 0) \to (1/2, 1/2), \phi}$ at a given temperature $T = 1 / \beta$ only depends on the two vortex energy barriers $F_{(0, 0) \to (1/2, 1/2), \phi}^{\mathrm{barrier}, \pm}$. The first vortex energy barrier $F_{(0, 0) \to (1/2, 1/2), \phi}^{\mathrm{barrier}, +}$ is plotted in Fig.~\ref{fig-6}(a) against the external field $\phi$ and shows a clear cusp (discontinuity in the first derivative) at a critical field $\phi_0^{+}$. Noting that the vortex energy barrier $F_{(0, 0) \to (1/2, 1/2), \phi}^{\mathrm{barrier}, +}$ is determined by the maximum of the vortex energy function $F_{0, 0, 1/2, 1/2, \phi}^{\mathrm{vortex}} (\varrho)$ in the vortex position $\varrho$ [see Eq.~(11) in the main text], it is illustrated in Fig.~\ref{fig-6}(b-d) that the critical field $\phi_0^{+}$ corresponds to a discontinuity in the vortex position $\varrho_0^{+}$ that maximizes the vortex energy function $F_{0, 0, 1/2, 1/2, \phi}^{\mathrm{vortex}} (\varrho)$. Analogously, the second vortex energy barrier $F_{(0, 0) \to (1/2, 1/2), \phi}^{\mathrm{barrier}, -}$ also has a cusp at another critical field $\phi_0^{-}$ corresponding to a discontinuity in the vortex position $\varrho_0^{-}$ that maximizes the vortex energy function $F_{1/2, 1/2, -1/2, -1/2, \phi}^{\mathrm{vortex}} (\varrho)$ [see Eq.~(11) in the main text].

For the specific location of the pinning site in Fig.~\ref{fig-5}(b), the two vortex energy barriers have identical critical fields: $\phi_0^{+} = \phi_0^{-}$ [see Fig.~\ref{fig-7}(a)]. However, if the pinning site is then moved inward or outward, the two critical fields $\phi_0^{\pm}$ shift in opposite directions and are generically different from each other [see Fig.~\ref{fig-7}(b-c)]. Consequently, the fractional magnetoresistance oscillations may develop a two-peak structure while retaining the same fractional periodicity (see Fig.~\ref{fig-7}). For generic disorder, we expect multiple features (not necessarily peaks) in the vortex energy barriers that are all replicated with the same periodicity. Importantly, while the precise shape and amplitude of the fractional oscillations thus depends on the specific disorder realization, the fractional periodicity itself is universal and only depends on the superfluid-density ratio $\gamma$ (see Fig.~\ref{fig-5}).

\clearpage

\end{widetext}


\begin{references}
\bibitem{Footnote-1} We remark that these quantities have dimensions of wave vector rather than current density.
\bibitem{Footnote-2} We note that the standard notation for this ratio is $\rho_{sp} / \rho_{s}$.
\bibitem{Footnote-3} We use $\xi / R_0 \approx 0.02$ but emphasize that its precise value is not important as $f_{nn} (\varrho)$ is only logarithmically divergent.
\bibitem{Footnote-4} While the pure spin vortices with $n_c = 0$ and $n_s = \pm 1$ have a smaller self energy for $\gamma < 1/3$, they do not couple to the bias current and hence do not contribute to the resistance.
\end{references}
\end{document}